\documentclass[sigconf]{acmart}
\acmConference{}{}{}

\renewcommand\footnotetextcopyrightpermission[1]{}

\usepackage{multirow}
\usepackage{algorithmic}
\usepackage{siunitx}
\usepackage{url}
\usepackage{subcaption}
\usepackage{enumitem}
\usepackage{listings}
\usepackage{xcolor}

\definecolor{codebg}{RGB}{242,242,235}  
\definecolor{codekw}{RGB}{237,2,140}   
\definecolor{codestr}{RGB}{120,  0,170}  
\definecolor{codeln}{gray}{0.55}      
\definecolor{codecm}{RGB}{0,128,0}
\lstdefinelanguage{PaperPython}{
  morekeywords={class,def,return,for,in,if,else,elif,while,break,continue,pass,
                True,False,None,import,from,as,with,try,except,finally,lambda,yield,
                assert,raise},
  keywordstyle=\color{codekw}\bfseries,
  comment=[l]\#,
  commentstyle=\color{codecm}\itshape,
  stringstyle=\color{codestr},
  morestring=[b]', morestring=[b]",
  sensitive=true,
}
\lstdefinestyle{NoHL}{
  basicstyle=\ttfamily\small,
  keywordstyle=,    
  commentstyle=,
  stringstyle=,
  identifierstyle=
}
\graphicspath{{Images/}}

\title{Code vs Serialized AST Inputs for LLM-Based Code Summarization: An Empirical Study}

\author{Shijia Dong}
\affiliation{
  \institution{University of Glasgow}
  \city{Glasgow}
  \country{United Kingdom}
}
\email{2810995d@student.gla.ac.uk}

\author{Haoruo Zhao}
\affiliation{
  \institution{University of Glasgow}
  \city{Glasgow}
  \country{United Kingdom}
}
\email{Haoruo.Zhao@glasgow.ac.uk}

\author{Paul Harvey}
\affiliation{
  \institution{University of Glasgow}
  \city{Glasgow}
  \country{United Kingdom}
}
\email{Paul.Harvey@glasgow.ac.uk}

\renewcommand{\shortauthors}{Dong, Zhao, and Harvey}

\begin{abstract}
Summarizing source code into natural language descriptions (code summarization) helps developers better understand program functionality and reduce the burden of software maintenance. Abstract Syntax Trees (ASTs), as opposed to source code, have been shown to improve summarization quality in traditional encoder–decoder-based code summarization models. However, most large language model (LLM)-based code summarization methods rely on raw code or only incorporate partial AST signals, meaning that the potential of complete AST representation has not been fully explored for LLMs. 

This paper presents AST(NIT), an AST augmentation and serialization method that preserves lexical details and encodes structural information into LLM-compatible sequences. Experiments with the LLaMA-3.1-8B model on the CodeXGLUE Python dataset show that the proposed serialized ASTs reduce the length of LLM inputs, require shorter training times, and achieve summarization quality comparable to existing approaches.

\end{abstract}


\begin{document}
\maketitle

\section{Introduction}
\label{sec:intro}
Code summarization aims to automatically generate concise natural language descriptions of source code, providing significant value for program understanding, software maintenance, and collaborative development~\cite{zhang2024review}. 

In existing encoder–decoder-based code summarization, explicitly incorporating structural and semantic information into input representations for the code summarization process provides richer context and improves performance~\cite{bansal2021project}. Among various input representations, Abstract Syntax Trees (ASTs)~\cite{baxter1998clone} are hierarchical, tree-based abstractions of source code that have been widely adopted and have demonstrated strong performance~\cite{hu2018deep,hu2020deep,leclair2019neural,tang2022ast}. While programming languages follow strict grammatical rules, models that process flat code tokens often struggle to capture the structural relationships inherent in code~\cite{lin2021improving}. ASTs address this limitation by explicitly representing code’s hierarchical structure, thereby providing richer context and enabling neural models to produce more accurate summaries. 

More recently, large language models (LLM)-based code summarization methods have demonstrated remarkable performance~\cite{fried2022incoder,li2023starcoder}. The successful adoption of ASTs in earlier encoder–decoder-based models suggests that leveraging complete ASTs may also benefit LLMs. However, in recent LLM-based code summarization, most works incorporate only partial structural signals from ASTs in the model input, such as data or control flow edges and tagged identifiers~\cite{ahmed2024automatic,lomshakov2024proconsul}. Such designs fail to preserve the complete structural context of the original tree, resulting in the loss of hierarchical relationships and control structures, potentially limiting the model’s ability to generate precise code summaries from the AST representation. One possible way to fully exploit the complete AST is to serialize the entire tree into a linear sequence, as in the Structure-Based Traversal (SBT) method proposed by Hu et al.~\cite{hu2018deep}, which has been shown to improve summarization quality in encoder–decoder settings. However, LLMs process inputs purely as sequences without explicit structural encoders, so it remains unclear whether the benefits of AST serialization extend to this setting. This leads to the following research question:

\textbf{RQ: Under LLM fine-tuning, can serialized ASTs achieve comparable or superior method-level summarization quality to code sequences?}

In this paper, \emph{serialized ASTs} refer to linearized representations of ASTs obtained through a tree-node traversal; \emph{code sequences} denote tokenized source code text without explicit structural information; \emph{Method-level summarization} focuses on generating single-sentence natural language descriptions for individual functions or methods. 

To address this question, we propose \textbf{AST(NIT)}, an AST augmentation and serialization method that preserves lexical details and encodes tree structure into LLM-compatible sequences. We systematically evaluate four input representations—AST(NIT), AST(Preorder), AST(SBT), and Code on the code summarization task using the CodeXGLUE (Python) subset~\cite{lu2021codexglue}, under identical fine-tuning and decoding settings. Experimental results show that, for method-level code summarization, serialized ASTs can achieve summarization quality comparable to code sequences when used as LLM inputs. Compared with the SBT method, AST(NIT) achieves comparable summary quality while reducing average input length by 28.6\% and total training time by 11.3\%, resulting in measurable efficiency improvements. In short, the contributions of this work are as follows:

\begin{itemize}
    \item AST(NIT), an AST augmentation and serialization method that preserves lexical details and encodes tree structure into LLM-compatible sequences;

    \item Empirical evaluation of complete serialized AST sequences as the sole input to LLMs for method-level code summarization;
    
    \item Systematic comparison of Code, AST(Preorder), AST(SBT), and AST(NIT) under an identical LLM fine-tuning and decoding setup. The results indicate that serialized ASTs  can achieve summarization quality comparable to code sequences, and that AST(NIT) provides measurable efficiency improvements over AST(SBT).
\end{itemize}

\section{Background \& Motivation}
\label{sec:background}
\subsection{Code Summarization}

Code summarization is the process of generating natural language descriptions for code~\cite{sun2024source}. An accurate summary helps developers quickly understand program intent, facilitating collaboration and software maintenance~\cite{zhang2024review}. The research in the code summarization field has evolved from early template-based and information retrieval methods~\cite{haiduc2010supporting,moreno2013automatic,eddy2013evaluating} to neural encoder-decoder-based models~\cite{iyer2016summarizing,fang2024esale,sun2024extractive,hu2018deep}. In recent years, LLMs have made rapid progress in code summarization~\cite{fried2022incoder,su2024distilled}. 

In addition to the continued evolution of model architectures, another prominent trend in this field is the incorporation of structural and semantic information~\cite{bansal2021project} such as ASTs~\cite{hu2018deep,allamanis2017learning}, data flow~\cite{guo2020graphcodebert}, and control flow graphs~\cite{ye2023cp} to enrich the input representation and provide models with richer context, thereby improving summary quality.

Typically, the granularity of existing code summarization approaches is at the operation, method, or class level~\cite{zhu2019automatic}. In this work, we focus on the method level, aiming to generate concise and accurate summaries for individual functions or methods.

\subsection{LLaMA\mbox{-}3.1}

LLMs are advanced neural models based on the transformer architecture~\cite{vaswani2017attention} and have achieved remarkable results in code understanding and generation tasks~\cite{wang2021codet5,roziere2023code}.

Among these models, we adopt \emph{LLaMA 3.1}, an open-source LLM released by Meta in 2024, due to its strong performance on code tasks and robust long-context modeling capabilities~\cite{dubey2024llama}. These features make it well suited for experiments that require processing large inputs, such as serialized AST sequences. \emph{LLaMA 3.1} is available in several parameter sizes (8B, 70B, 405B); we select the
\emph{LLaMA\mbox{-}3.1\mbox{-}8B}\footnote{\url{https://huggingface.co/meta-llama/Llama-3.1-8B-Instruct}} as it provides a good balance between performance and computational feasibility for fine-tuning on a single GPU (see Section~\ref{sec:experimental} for details).

\subsection{Abstract Syntax Tree (AST)}

An AST~\cite{baxter1998clone} is a hierarchical, tree-based representation of the abstract syntactic structure of source code, explicitly encoding both syntactic and structural information. AST representations, when used as model inputs, have demonstrated notable performance improvements across various code-related tasks, such as code summarization~\cite{leclair2019neural}, code search~\cite{hu2022tackling}, code clone detection~\cite{zhang2019novel}. In a typical AST, terminal (leaf) nodes represent variables and types, while non-terminal (internal) nodes denote syntactic constructs such as loops, expressions, or declarations~\cite{tang2022ast}. 

\begin{figure*}[t]
  \centering
  \begin{subfigure}[t]{0.5\linewidth}
    \centering
    \includegraphics[width=\linewidth]{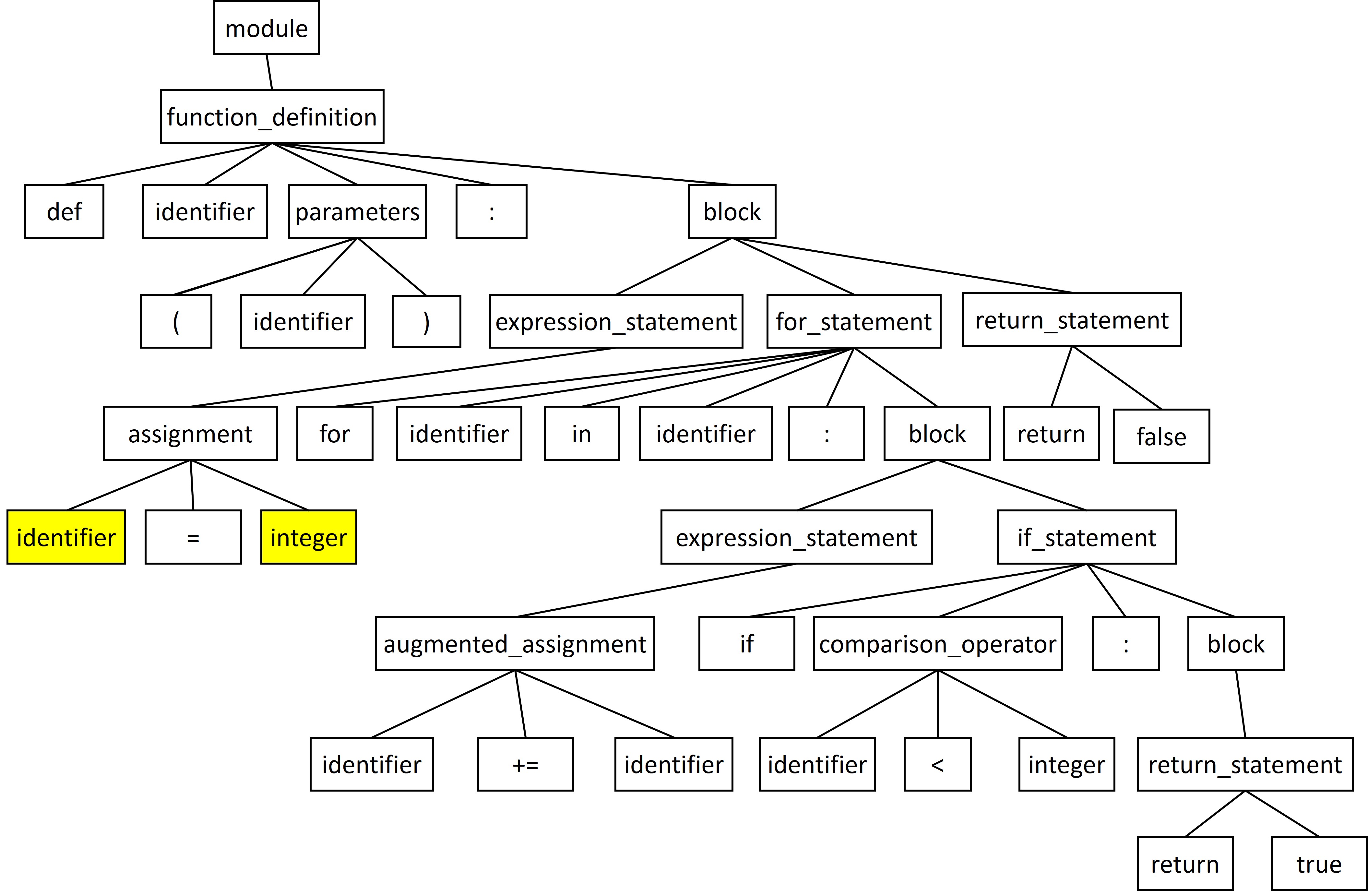}
    \caption{Raw AST from Tree-sitter.}\label{fig:ast_raw}
  \end{subfigure}\hfill
  \begin{subfigure}[t]{0.5\linewidth}
    \centering
    \includegraphics[width=\linewidth]{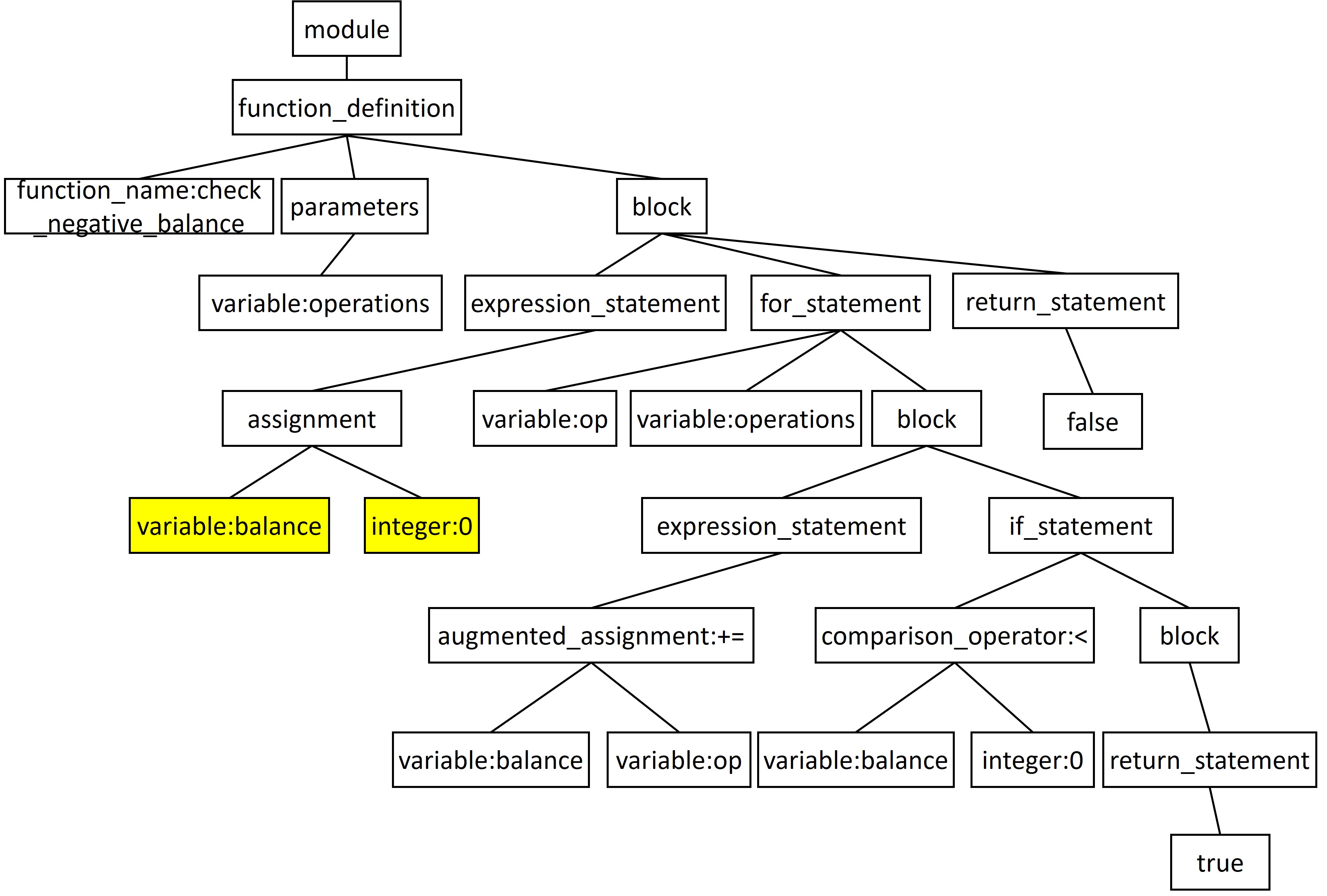}
    \caption{AST after lexical injection and structural normalization}\label{fig:ast_aug}
  \end{subfigure}
  \caption{Side-by-side comparison of (a) the raw AST and (b) the augmented AST for Listing~1.}
   \Description{Two side-by-side tree diagrams for the same Python function. The left diagram shows the raw abstract syntax tree produced by Tree-sitter. The right diagram shows an augmented tree after lexical injection and structural normalization, resulting in a smaller tree while preserving the original structure.}
  \label{fig:ast_compare}
\end{figure*}

\begin{figure*}[t]
    \centering
    \includegraphics[width=0.95\linewidth]{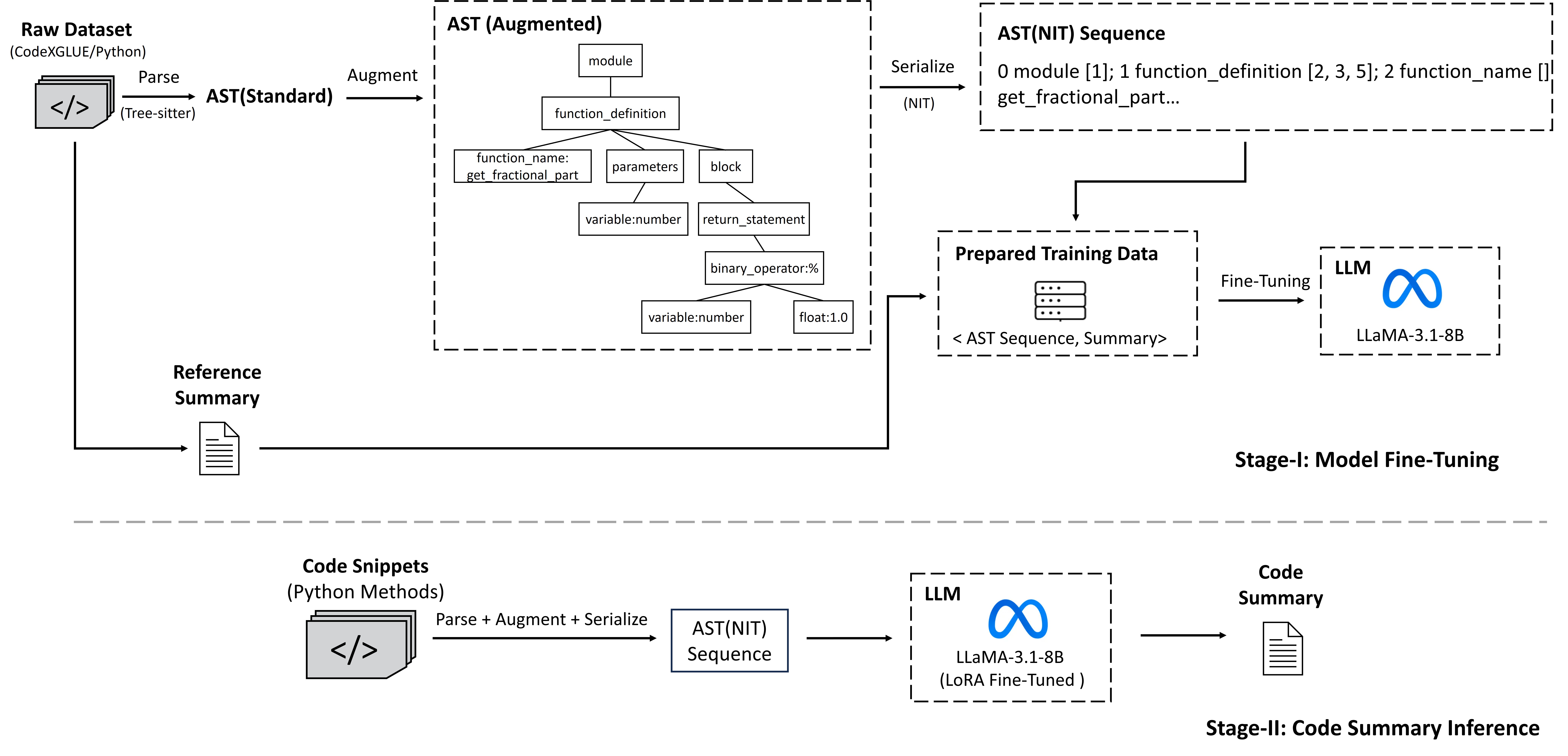} 
    \caption{The overall workflow of the proposed method AST(NIT) for code summarization with serialized AST inputs.}
    \Description{Workflow of AST(NIT) code summarization with two stages: (I) training, where CodeXGLUE Python code is parsed by Tree-sitter, the AST is augmented and serialized into an AST(NIT) sequence, and sequence-summary pairs are used to fine-tune LLaMA-3.1-8B; (II) inference, where new code is converted into an AST(NIT) sequence and passed to the fine-tuned model to generate a summary.}
    \label{fig:overview}
\end{figure*}

\subsection{Motivation}

Inspired by the strong performance of ASTs in encoder–decoder-based models for code summarization, recent LLM-based approaches have increasingly explored the potential of ASTs. However, most existing LLM studies utilize only partial structural signals from ASTs, such as data flow or tagged identifiers, incorporated as prompt augmentations by appending these structural hints to the text prompt~\cite{ahmed2024automatic,lomshakov2024proconsul}. Compared with using the full AST as input, this partial-structure design overlooks the complete structural context of the original tree, causing the hierarchical relationships and control structures within the AST to be lost. This limits the model’s ability to accurately capture code semantics (functional meaning and logical intent) from the AST representation and to generate precise code summaries.

Given that SBT methods~\cite{hu2018deep,leclair2019neural} have shown that AST serialization can effectively support code summarization within encoder–decoder frameworks, a potential solution for leveraging complete AST in LLM-based code summarization is to use a serialized AST as an independent input to the model. However, there are two main challenges that need to be overcome:

\begin{lstlisting}[language=Python,caption={Check negative balance}]
def check_negative_balance(operations):
    balance = 0
    for op in operations:
        balance += op
        if balance < 0:
            return True
    return False
\end{lstlisting}

(1) \textbf{Loss of lexical detail.}
Prior work has shown that identifiers carry critical lexical details from the original code, strongly influence model performance on code summarization tasks~\cite{ahmed2022few}. However, in standard ASTs, these identifiers are often abstracted away, as terminal nodes typically record only the type without the concrete values~\cite{tang2022ast}. In Listing 1, for the statement \emph{"balance = 0"} (line 2), the nodes corresponding to \emph{"balance"} and \emph{"0"} are represented only as \emph{"identifier"} and \emph{"integer"} (Fig.~\ref{fig:ast_raw}). 

(2) \textbf{Structural mismatch.} Without a structure-aware encoder, the key challenge is to serialize the AST while preserving its original hierarchical relationships (i.e., the parent–child and nested node structure that reflects how statements and expressions are organized in the code), ensuring LLMs can effectively capture and utilize this structural information to generate accurate summaries.

\section{Methods}
\label{sec:methods}
To address the challenges outlined in Section~\ref{sec:background}, we propose AST(NIT), an AST augmentation and serialization method that preserves lexical details and encodes tree structure into LLM-compatible sequences. 

Our workflow (Fig.~\ref{fig:overview}) consists of two stages: (i) \textbf{model fine-tuning}, which includes parsing, AST augmentation, AST serialization, and fine-tuning, and (ii) \textbf{code summary inference}. The first three steps: parsing, AST augmentation, and AST serialization together constitute AST(NIT). We describe the implementation of AST(NIT) below, while the fine-tuning configuration is detailed in Section~\ref{sec:experimental}. The code summary inference stage follows the trained model’s decoding setup, also described in Section~\ref{sec:experimental}.

\subsection{AST Augmentation}
In this work, we consider Python code as prior work~\cite{tang2022ast} reports that
AST encoding is more effective for Python, as its corresponding ASTs are relatively shorter and thus allow structural information to contribute more significantly than in Java. We use the open-source Python project \emph{Tree-sitter}\footnote{\url{https://tree-sitter.github.io/tree-sitter}} to parse code snippets into ASTs. However, as discussed in Section~\ref{sec:background}, standard ASTs generated by such parsers do not retain all lexical details from the source code. Moreover, standard ASTs are typically large (see Figure~\ref{fig:ast_raw}) and contain numerous unnamed nodes~\cite{beckmann2024supporting}, such as commas and parentheses. For some complex methods, serializing the corresponding ASTs can produce long sequences that may approach or exceed the context window of an LLM. In addition, longer inputs increase computational cost and can diffuse the model’s attention over less informative tokens, ultimately degrading the quality of generated code summaries. To address these issues, we modify the parsed ASTs as follows:

(1) \textbf{Lexical Injection.} We explicitly inject lexical information into terminal nodes by appending identifier names and embedding numeric or string literals. In addition, we refine certain node types to better reflect their semantic roles: for example, an \emph{"identifier"} within a function definition is relabeled as \emph{"function\_name"}, while identifiers in assignment expressions are relabeled as \emph{"variable"}. After modification, the nodes corresponding to \emph{"balance"} and \emph{"0"} become \emph{"variable:balance"} and \emph{"integer:0"}, as highlighted in Figures~\ref{fig:ast_raw} and~\ref{fig:ast_aug}.

(2) \textbf{Structural Normalization.} We remove most unnamed nodes unless they carry clear semantic value (such as colons in slice expressions). For operator nodes, we embed the actual operator as its value (e.g., \emph{"binary\_operator:+"}) rather than creating a separate child node. This effectively reduces the size of the tree and eliminates irrelevant noise. For the example shown in Figures~\ref{fig:ast_raw} and~\ref{fig:ast_aug}, the number of AST nodes decreases from \textbf{41} in the raw tree to \textbf{27} after structural normalization.

This modification augments the AST with lexical details for code summarization, while reducing overall tree size and preserving its original structure and depth, making it suitable as input to LLMs. For comparison, we also evaluate the raw AST without any lexical injection or structural normalization in Section~\ref{sec:experimental} (see AST(Preorder)).

\subsection{AST Serialization}

To make the augmented AST compatible with sequence-based LLMs, we serialize it into a flat token sequence using a \textbf{N}ode-\textbf{I}ndex \textbf{T}raversal (\textbf{NIT}). We refer to the final linear representation as \textbf{AST(NIT)}, emphasizing that it includes both augmentation and serialization.

\textbf{Traversal procedure:}
Starting at the root, we perform a Depth-First Search (DFS) based preorder traversal~\cite{tarjan1972depth} and assign each visited node a globally unique integer identifier (ID) in order of visitation. Each node is recorded as fixed-field tuple and appended into the token sequence, with individual nodes separated by semicolons (\texttt{;}). The tuple comprises the following fields:
\begin{itemize}
    \item \textbf{ID}: the unique index of the node;
    \item \textbf{Type}: the node type (e.g., \emph{"function\_definition"}, \emph{"call"});
    \item \textbf{Value} (optional): concrete lexical content, if present (e.g., identifier names, literals, embed operators);
    \item \textbf{Children} (optional): the list of child IDs in visitation order, if any.
\end{itemize}

We use a DFS-based preorder traversal because it follows the nested structure of code, exploring each execution path to completion before backtracking. For each node, we record its attributes in a fixed field order, making the structural information explicit in the serialized sequence. As a result, the entire AST can be encoded as a flat sequence, which can then be used as a direct input to LLMs.

\section{Experimental Evaluation}
\label{sec:experimental}
To systematically evaluate serialized AST representations for code summarization, we first introduce the experimental setup and evaluation metrics, and then describe the four input representations in this section: Code, AST(Preorder), AST(SBT), and AST(NIT). Quantitative results and qualitative analysis are presented in Section~\ref{sec:results}.

\subsection{Experimental Setup}
\textbf{Dataset.}
We use the Python subset of the code summarization task in CodeXGLUE~\cite{lu2021codexglue}, a widely used benchmark dataset for program understanding. We first randomly sample approximately 50{,}000 methods for training, 5{,}000 for validation, and 5{,}000 for testing. We then apply the filtering procedure to the existing records in the dataset adopted in prior work~\cite{hu2018deep,tang2022ast}: (i) remove methods whose reference summary contain fewer than four words; 
(ii) exclude constructors, property accessors, and test cases, as their summaries are typically trivial for the model to generate and may artificially inflate performance estimates~\cite{hu2018deep};
(iii) discard duplicate samples; 
(iv) if a summary contains multiple sentences, retain only the first sentence; and 
(v) remove samples whose AST cannot be successfully parsed. 
After cleaning, the final splits contain 30{,}227 training, 2{,}771 validation, and 3{,}097 test instances~(Table~\ref{tab:dataset}).

\begin{table}[h]
\centering
\setlength{\tabcolsep}{3pt}
\caption{Dataset statistics (token counts measured with the LLaMA-3.1 tokenizer).}
\label{tab:dataset}
\begin{tabular}{r r ccc ccc}
\toprule
\multirow{2}{*}{Split} & \multirow{2}{*}{Count} &
\multicolumn{3}{c}{Code length (tokens)} &
\multicolumn{3}{c}{Summary length (tokens)} \\
\cmidrule(lr){3-5}\cmidrule(lr){6-8} 
 & & Min & Mean & Max & Min & Mean & Max \\
\midrule
Train & 30227 & 16 & 85.97 & 474 & 4 & 9.47 & 38 \\
Valid &  2771 & 17 & 84.15 & 395 & 4 & 9.43 & 31 \\
Test  &  3097 & 16 & 85.20 & 433 & 4 & 9.33 & 31 \\
\bottomrule
\end{tabular}
\end{table}

\begin{table*}[h]
\centering
\caption{Example input representations for the same function, from Listing~1 (truncated, * denotes our method).}
\label{tab:input_examples}
\begin{tabular}{lp{13cm}}
\toprule
Representation & Example \\
\midrule
AST(NIT)* & \lstinline|0 module [1]; 1 function_definition [2,3,5]; 2 function_name [] check_negative_balance; 3 parameters [4]; 4 identifier [] operations; 5 block [6,10,25]; 6 expression_statement [7]; 7 assignment [8,9]; 8 identifier [] balance; 9 integer [] 0; ...| \\
AST(Preorder) & \lstinline|module function_definition identifier parameters identifier block expression_statement assignment identifier ...| \\
AST(SBT) & \lstinline|(module (function_definition (function_name_check_negative_balance) function_name_check_negative_balance (parameters (identifier_operations) identifier_operations )parameters (block (expression_statement (assignment (identifier_balance) identifier_balance (integer_0) integer_0 )assignment )expression_statement ...| \\
Code & \lstinline[style=NoHL]|def check_negative_balance(operations): balance = 0 ...| \\

\bottomrule
\end{tabular}
\\[3pt]

\end{table*}

\textbf{Model.}
To ensure a fair comparison across input representations, we fine-tune the \emph{LLaMA-3.1-8B} model using the above datasets, rather than relying solely on its pre-trained parameters, thereby reducing bias from pre-training and prompt sensitivity. All experiments use the \emph{Meta-Llama-3.1-8B-Instruct}\footnote{\url{https://huggingface.co/meta-llama/Llama-3.1-8B-Instruct}} checkpoint with 4-bit quantization via the \emph{BitsAndBytes}\footnote{\url{https://github.com/bitsandbytes-foundation/bitsandbytes}} library, and the same configuration is applied for all input representations.

\textbf{Training Details.} We apply Parameter-Efficient Fine-Tuning~\cite{ding2023parameter} with \emph{LoRA}~\cite{hu2022lora}, implemented using the \emph{unsloth}\footnote{\url{https://unsloth.ai}} library which substantially reduces memory and time cost. \emph{LoRA} is applied with rank $r = 16$, $\alpha = 16$, and dropout = $0.05$ to the projection layers, combined with gradient checkpointing for memory efficiency. Across all input representations, we use the \textbf{same} hyper-parameters: learning rate = $5\times 10^{-5}$, warm-up ratio = $0.05$, weight decay = $0.01$, and max gradient norm = $1.0$. We train for three epochs with a context window of 5{,}000 tokens for all conditions, using the AdamW (8-bit) optimiser~\cite{loshchilov2017decoupled} and mixed-precision. Since the average input length varies across the four input representations—Code, AST(Preorder), AST(SBT), and AST(NIT), we adopt token-based batching targeting $\approx 50{,}000$ tokens per update to reduce gradient variance. All checkpoints are retained during training, and the checkpoint with the highest validation BLEU-4~\cite{papineni2002bleu} is selected for test evaluation. Inference uses deterministic decoding (beam = 4, length penalty = 0.6, max\_new\_tokens = 64). All input types share the same zero-shot style template.

\textbf{Hardware.} All models are trained on a single NVIDIA A6000 GPU (48GB).

\subsection{Evaluation Metrics}
We evaluate the generated summaries on the test set using four widely adopted metrics for code summarization:  
(1) \textbf{BLEU-4}~\cite{papineni2002bleu} measures the precision of overlapping $n$-grams between a generated summary and its reference. Following prior work~\cite{hu2018deep}, we set $n=4$ to balance local phrase accuracy with overall fluency. It captures how many contiguous sequences of up to four tokens appear in both texts. 
(2) \textbf{METEOR}~\cite{banerjee2005meteor} computes similarity through unigram alignment between the generated and reference summaries. Unlike BLEU-4, METEOR allows flexible matching based on exact forms, stemming, and synonyms~\cite{lin2021improving}, making it more sensitive to linguistic variation and synonym usage.  
(3) \textbf{ROUGE-L}~\cite{lin2004rouge} evaluates sentence-level similarity by identifying the longest common subsequence between a generated and a reference summary. By focusing on the longest in-sequence matches, ROUGE-L rewards summaries that preserve the overall order of key words or phrases, rather than isolated $n$-gram matches.  
(4) \textbf{BERTScore}~\cite{zhang2019bertscore} moves beyond lexical overlap and measures semantic similarity by leveraging contextualised embeddings from a variant of BERT~\cite{sun2024source}. Each token in the generated summary is aligned to its most similar reference token using embedding-based cosine similarity, and overall precision, recall, and F1 scores are then computed. This allows BERTScore to capture semantic equivalence even when surface forms differ.

\subsection{Baselines}
We compare the performance of the following four input representations on the code summarization task. Models trained with different input representations share \textbf{identical} hyperparameters and training/inference settings. Table~\ref{tab:input_examples} shows encodings of Listing 1 in each representation.

\begin{itemize}
  \item[(1)] \textbf{AST(NIT) (our method).} An AST sequence obtained by first parsing the source code with \emph{Tree-sitter}, then applying Lexical Injection and Structural Normalization to augment the tree, and finally serializing the augmented AST using our proposed Node-Index Traversal (NIT) (see Section~\ref{sec:methods})
  
  \item[(2)] \textbf{AST(Preorder).} An AST sequence obtained by parsing the source code with \emph{Tree-sitter} and serializing the unmodified tree via preorder traversal without lexical injection or structural normalization. AST(Preorder) is a purely structural representation: it retains only AST node types and excludes identifier names and literal values.

  \item[(3)]  \textbf{AST(SBT).} An AST sequence obtained by parsing with \emph{Tree-sitter} and serialized via Structure-Based Traversal (SBT)~\cite{hu2018deep}. SBT encodes tree structure using bracket-style markers and injects concrete lexical values at terminal nodes. Although originally demonstrated on Java, we apply the same serialization strategy to Python to ensure a fair comparison. AST(SBT) thus preserves identifier-level lexical information and does not require the raw source code as a separate input.

  \item[(4)] \textbf{Code.} The source code token sequence. As a lexical representation that does not explicitly encode structural information, Code constitutes the most direct input format for code-related tasks. We use it as a baseline to compare the effectiveness of explicit AST-based input representations.

\end{itemize}

\section{Results}
\label{sec:results}
To address the research question stated in Section~\ref{sec:intro}, we evaluate the four input representations from two perspectives: (i) summary quality, assessed by BLEU-4, METEOR, ROUGE-L, and BERTScore under identical fine-tuning and decoding settings; and (ii) efficiency, measured by average input length, total trained tokens, training time, and peak memory usage. Qualitative analysis and discussion are also presented in this section.

\begin{table}[h]
\setlength{\tabcolsep}{2pt}
\centering
\caption{Comparison of code summarization performance on BLEU, METEOR, ROUGE-L, and BERTScore metrics for different input representations (* denotes our method).}
\label{tab:input_evaluation}
\begin{tabular}{ccccccc}
\toprule
\multirow{2}{*}{Input} &
\multirow{2}{*}{BLEU-4} &
\multirow{2}{*}{METEOR} &
\multirow{2}{*}{ROUGE-L} &
\multicolumn{3}{c}{BERTScore} \\
\cmidrule(lr){5-7}
 &  &  &  & Precision & Recall & F1 \\
\midrule
AST(NIT)*     & 23.07 & 0.39 & 0.48 & 0.93 & 0.91 & 0.92 \\
AST(Preorder) & 11.75 & 0.19 & 0.25 & 0.89 & 0.88 & 0.89 \\
AST(SBT)      & 23.22 & 0.40 & 0.48 & 0.93 & 0.91 & 0.92 \\
Code          & 23.48 & 0.39 & 0.49 & 0.93 & 0.91 & 0.92 \\
\bottomrule
\end{tabular}
\end{table}

\subsection{Code vs AST(NIT) vs AST(Preorder)}  Table~\ref{tab:input_evaluation} shows that, across all evaluation metrics, Code and AST(NIT) achieve near-identical results. This indicates that, under LLM fine-tuning, using either code sequences or a structured representation like AST(NIT) as input yields comparable code summarization performance. However, AST(Preorder) substantially underperforms (e.g., BLEU-4: 11.75). This is mainly because the absence of lexical details in AST(Preorder): user-defined identifiers and literals are collapsed into generic node types, leading to semantic loss. By contrast, Code retains lexical cues natively, and AST(NIT) reintroduces them via lexical injection. These observations indicate that lexical information is essential for code summarization, align with prior findings~\cite{ahmed2022few}, and indicate the effectiveness of the AST-augmentation design used in AST(NIT). 

\subsection{AST(SBT) vs AST(NIT)} Table~\ref{tab:input_evaluation} shows that our proposed method, AST(NIT), achieves summarization quality comparable to the AST(SBT) across all evaluation metrics. From the efficiency perspective (Table~\ref{tab:training}), AST(NIT) reduces the average input length by approximately 28.6\% compared to AST(SBT), and shortens total training time by 11.3\% , with similar peak memory usage. This efficiency gains come from AST(NIT)’s compact fixed-field node tuple and child-ID list design, which avoids the repeated type markers and bracket nesting of AST(SBT). In summary, our method achieves summarization performance comparable to AST(SBT) while using shorter input sequences and has lower training cost.

\begin{table}[h]
\setlength{\tabcolsep}{2pt}
\centering
\caption{Training statistics across different input representations (* denotes our method).}
\label{tab:training}
\begin{tabular}{ccccc}
\toprule
Input & Avg Length & Total Trained & Training & Peak Memory \\
Representation & (tokens) & Tokens (M) & Time (h) & Usage (GB) \\
\midrule
AST(NIT)*     & 470.92 & 14.23M & 11.81 & 11.50 \\
AST(Preorder) & 133.94 &  4.05M &  3.57 &  8.80 \\
AST(SBT)      & 659.17 & 19.90M & 13.32 & 11.50 \\
Code          & 117.82 &  3.56M &  3.04 &  7.61 \\
\bottomrule
\end{tabular}
\end{table}

\subsection{Qualitative Analysis} 
We present two representative examples to qualitatively assess the semantic accuracy of summaries generated from different input representations.

In Listing 2, summaries from Code, AST(NIT), and AST(SBT) all capture the intended file deletion operation, with minor lexical differences such as \emph{"lock file"} or \emph{"temporary file"}, as shown in Table~\ref{tab:listing2}. In contrast, AST(Preorder) incorrectly describes the action as \emph{"removing a directory"}, missing the file-specific context.

\begin{lstlisting}[language=Python,caption={cleanup}]
def cleanup(self):
    if os.path.exists(self.path):
        os.remove(self.path)
\end{lstlisting}

\begin{table}[h]
\setlength{\tabcolsep}{3pt} 
\centering
\caption{Generated Summaries for Listing 2 Across Input Representations}
\label{tab:listing2}
\begin{tabular}{ll}
\toprule
Reference & Clean up files in the specified path.\\
\midrule
AST(NIT)*    & Removes the temporary file.\\
AST(SBT)    & Removes the temporary file.\\
AST(Preorder) & Removes the directory. \\
Code        & Remove the lock file.\\
\bottomrule
\end{tabular}
\end{table}

In Listing 3, both Code and lexical-injected AST inputs (AST(NIT), AST(SBT)) accurately describe the creation of a virtual environment, as shown in Table~\ref{tab:listing3}. However, AST(Preorder) fails to capture this intent and instead generates a summary about creating a file instance.

\begin{lstlisting}[caption={create}]
def create(env_dir, system_site_packages=False, clear=False, symlinks=False, with_pip=False, prompt=None):
    builder = ExtendedEnvBuilder(
        system_site_packages=system_site_packages,
        clear=clear,
        symlinks=symlinks,
        with_pip=with_pip,
        prompt=prompt
    )
    builder.create(env_dir)
    return builder.context
\end{lstlisting}

\begin{table}[h]
\setlength{\tabcolsep}{2pt} 
\centering
\caption{Generated Summaries for Listing 3 Across Input Representations}
\label{tab:listing3}
\begin{tabular}{ll}
\toprule
Reference & Create a virtual environment in a directory. \\
\midrule
AST(NIT)*    & Create a virtual environment in the directory. \\
AST(SBT)    & Create a virtual env in the given directory. \\
AST(Preorder) & Create an instance of file. \\
Code        & Create a virtual environment in the given directory. \\
\bottomrule
\end{tabular}
\end{table}

In these two examples, when lexical information is preserved, either natively in Code or via lexical injection in AST(NIT) and AST(SBT), the generated summaries are generally more specific and better capture the intended functionality. In contrast, purely structural inputs such as AST(Preorder) often result in information loss and less precise outputs, consistent with Table~\ref{tab:input_evaluation}. Meanwhile, AST(NIT) produces code summaries comparable to those from Code in both accuracy and detail. 

\subsection{Discussion}
 
Together the results from Table~\ref{tab:input_evaluation} with qualitative analysis, we find that while serialized ASTs (e.g., SBT) once offered clear benefits in encoder–decoder model, our empirical evaluation suggests that, under LLM fine-tuning, serialized ASTs achieve summary quality comparable to code sequences.

This shift can be explained by two possible situations. First, emergent abilities from large-scale pretraining enable LLMs to internalize structural and semantic information directly from code without requiring explicit AST signals. Second, because modern LLMs are pretrained on massive corpus that likely include human-curated datasets, or close variants, used in current benchmarks, the evaluation setting differs fundamentally from that of traditional encoder–decoder models. In our fine-tuning scenarios, it is no longer possible to guarantee a strict separation between training and test sets, meaning that the model may already have encountered the benchmark data or similar examples during pretraining. This overlap makes it harder to obtain clear insights into the comparative evaluation of AST versus code inputs. Our findings do not prevent the possibility that AST representations remain valuable in specialized tasks or data-scarce settings, where explicit structural signals may still provide complementary benefits.

\section{Related Work}
\subsection{AST Serialization Techniques}

To enable sequence-based neural models to process ASTs, researchers commonly employ traversal methods to serialize tree structures into sequential representations.

Classic traversal strategies, such as preorder and postorder, generate serialized sequences by recording node types and visitation order, but typically do not support lossless reconstruction of the original AST~\cite{zhang2020retrieval}. To address the limitations of classic traversals, SBT~\cite{hu2018deep,hu2020deep,leclair2019neural} introduces bracket-style markers to enable unambiguous reconstruction of the original tree. Similarly, multi-sequence approaches, such as providing both root and leaf node sequences or combining preorder traversal sequences with parent node sequences~\cite{lin2021semantic,qiu2024software}, also support reconstructing the original AST. However, these methods typically result in much longer and more redundant sequences, which increases computational cost. Besides, some studies employ path-based methods, such as Code2Seq and PathMiner~\cite{alon2018code2seq, kovalenko2019pathminer}, which serialize ASTs as collections of root-to-leaf paths or subtrees. These methods also preserve the structural integrity of the AST. However, the number of path combinations is huge, which leads to input expansion while only sampling selection is difficult to cover the global structure.

While tree and graph-based neural models~\cite{wan2018improving,lin2021improving} capture explicit hierarchy, sequence-based AST serializations still have the advantage in LLM pipelines due to their compatibility with transformer architectures and direct support for efficient, end-to-end processing.

\subsection{LLMs for Code Summarization}

Recent LLM-based code summarization studies can be grouped by their input:

\begin{itemize}
\item[(1)] Some models use only the code sequence as input. For example, StarCoder~\cite{li2023starcoder}, InCoder~\cite{fried2022incoder}, and CodeLlama~\cite{roziere2023code} generate summaries directly from tokenized source code, without adding structural information.

\item[(2)] To provide LLMs with richer context, thereby improving summary quality. Some other models augment the input with additional context or structure. These methods add information such as function signatures, comments, documentation, or file paths, and may include structural signals such as call or dependency graphs, data-flow information. Examples include PROCONSUL~\cite{lomshakov2024proconsul}, which uses call graphs, and Ahmed et~al.~\cite{ahmed2024automatic}, which supplement few-shot prompts with repository path and data-flow details.
\end{itemize}

Research in the second category is more related to our work, as ASTs can be seen as structure-augmented input. This motivates us to examine whether serialized AST representations may achieve comparable or superior summarization quality over code input with LLMs in a fine-tuning context.

\section{Conclusion and Future Work}
This paper investigates whether serialized ASTs can achieve comparable or superior method-level summarization quality to code sequences. To this end, we propose AST(NIT), an AST augmentation and serialization technique that preserves lexical details and serializes the tree as a compact sequence via node-index traversal. Using the \emph{LLaMA-3.1-8B}, we systematically compare AST(NIT) with two established AST serializations (SBT and preorder) as well as with code sequences, evaluating both summary quality and efficiency. Our results show that serialized ASTs can achieve summary quality comparable to code sequences; moreover, compared to SBT, AST(NIT) significantly reduces average input length and training time. 

Future work will examine whether these findings generalize across other programming languages, datasets, and LLM variants. We also plan to explore the applicability of AST(NIT) to other software engineering tasks where structural information may be more directly relevant, such as semantic code clustering and software modularization. In addition, human evaluation may be considered to complement automatic metrics.

\bibliographystyle{ACM-Reference-Format}
\bibliography{report}

@article{lu2021codexglue,
  title={Codexglue: A machine learning benchmark dataset for code understanding and generation},
  author={Lu, Shuai and Guo, Daya and Ren, Shuo and Huang, Junjie and Svyatkovskiy, Alexey and Blanco, Ambrosio and Clement, Colin and Drain, Dawn and Jiang, Daxin and Tang, Duyu and others},
  journal={arXiv preprint arXiv:2102.04664},
  year={2021}
}

@inproceedings{hu2018deep,
  title={Deep code comment generation},
  author={Hu, Xing and Li, Ge and Xia, Xin and Lo, David and Jin, Zhi},
  booktitle={Proceedings of the 26th conference on program comprehension},
  pages={200--210},
  year={2018}
}

@inproceedings{tang2022ast,
  title={Ast-trans: Code summarization with efficient tree-structured attention},
  author={Tang, Ze and Shen, Xiaoyu and Li, Chuanyi and Ge, Jidong and Huang, Liguo and Zhu, Zhelin and Luo, Bin},
  booktitle={Proceedings of the 44th International Conference on Software Engineering},
  pages={150--162},
  year={2022}
}

@article{hu2022lora,
  title={Lora: Low-rank adaptation of large language models.},
  author={Hu, Edward J and Shen, Yelong and Wallis, Phillip and Allen-Zhu, Zeyuan and Li, Yuanzhi and Wang, Shean and Wang, Lu and Chen, Weizhu and others},
  journal={ICLR},
  volume={1},
  number={2},
  pages={3},
  year={2022}
}

@inproceedings{baxter1998clone,
  title={Clone detection using abstract syntax trees},
  author={Baxter, Ira D and Yahin, Andrew and Moura, Leonardo and Sant'Anna, Marcelo and Bier, Lorraine},
  booktitle={Proceedings. International Conference on Software Maintenance (Cat. No. 98CB36272)},
  pages={368--377},
  year={1998},
  organization={IEEE}
}

@article{dubey2024llama,
  title={The llama 3 herd of models},
  author={Dubey, Abhimanyu and Jauhri, Abhinav and Pandey, Abhinav and Kadian, Abhishek and Al-Dahle, Ahmad and Letman, Aiesha and Mathur, Akhil and Schelten, Alan and Yang, Amy and Fan, Angela and others},
  journal={arXiv e-prints},
  pages={arXiv--2407},
  year={2024}
}

@article{loshchilov2017decoupled,
  title={Decoupled weight decay regularization},
  author={Loshchilov, Ilya and Hutter, Frank},
  journal={arXiv preprint arXiv:1711.05101},
  year={2017}
}

@inproceedings{banerjee2005meteor,
  title={METEOR: An automatic metric for MT evaluation with improved correlation with human judgments},
  author={Banerjee, Satanjeev and Lavie, Alon},
  booktitle={Proceedings of the acl workshop on intrinsic and extrinsic evaluation measures for machine translation and/or summarization},
  pages={65--72},
  year={2005}
}

@inproceedings{lin2021improving,
  title={Improving code summarization with block-wise abstract syntax tree splitting},
  author={Lin, Chen and Ouyang, Zhichao and Zhuang, Junqing and Chen, Jianqiang and Li, Hui and Wu, Rongxin},
  booktitle={2021 IEEE/ACM 29th International Conference on Program Comprehension (ICPC)},
  pages={184--195},
  year={2021},
  organization={IEEE}
}

@inproceedings{papineni2002bleu,
  title={Bleu: a method for automatic evaluation of machine translation},
  author={Papineni, Kishore and Roukos, Salim and Ward, Todd and Zhu, Wei-Jing},
  booktitle={Proceedings of the 40th annual meeting of the Association for Computational Linguistics},
  pages={311--318},
  year={2002}
}

@inproceedings{lin2004rouge,
  title={Rouge: A package for automatic evaluation of summaries},
  author={Lin, Chin-Yew},
  booktitle={Text summarization branches out},
  pages={74--81},
  year={2004}
}

@article{sun2024source,
  title={Source code summarization in the era of large language models},
  author={Sun, Weisong and Miao, Yun and Li, Yuekang and Zhang, Hongyu and Fang, Chunrong and Liu, Yi and Deng, Gelei and Liu, Yang and Chen, Zhenyu},
  journal={arXiv preprint arXiv:2407.07959},
  year={2024}
}

@article{zhang2019bertscore,
  title={Bertscore: Evaluating text generation with bert},
  author={Zhang, Tianyi and Kishore, Varsha and Wu, Felix and Weinberger, Kilian Q and Artzi, Yoav},
  journal={arXiv preprint arXiv:1904.09675},
  year={2019}
}

@article{ding2023parameter,
  title={Parameter-efficient fine-tuning of large-scale pre-trained language models},
  author={Ding, Ning and Qin, Yujia and Yang, Guang and Wei, Fuchao and Yang, Zonghan and Su, Yusheng and Hu, Shengding and Chen, Yulin and Chan, Chi-Min and Chen, Weize and others},
  journal={Nature Machine Intelligence},
  volume={5},
  number={3},
  pages={220--235},
  year={2023},
  publisher={Nature Publishing Group UK London}
}

@article{tarjan1972depth,
  title={Depth-first search and linear graph algorithms},
  author={Tarjan, Robert},
  journal={SIAM journal on computing},
  volume={1},
  number={2},
  pages={146--160},
  year={1972},
  publisher={SIAM}
}

@inproceedings{beckmann2024supporting,
  title={Supporting Construction of Domain-Specific Representations in Textual Source Code},
  author={Beckmann, Tom and Reppien, Jan and Lincke, Jens and Hirschfeld, Robert},
  booktitle={Proceedings of the 3rd ACM SIGPLAN International Workshop on Programming Abstractions and Interactive Notations, Tools, and Environments},
  pages={17--28},
  year={2024}
}

@article{hu2020deep,
  title={Deep code comment generation with hybrid lexical and syntactical information},
  author={Hu, Xing and Li, Ge and Xia, Xin and Lo, David and Jin, Zhi},
  journal={Empirical Software Engineering},
  volume={25},
  number={3},
  pages={2179--2217},
  year={2020},
  publisher={Springer}
}

@inproceedings{zhang2019novel,
  title={A novel neural source code representation based on abstract syntax tree},
  author={Zhang, Jian and Wang, Xu and Zhang, Hongyu and Sun, Hailong and Wang, Kaixuan and Liu, Xudong},
  booktitle={2019 IEEE/ACM 41st International Conference on Software Engineering (ICSE)},
  pages={783--794},
  year={2019},
  organization={IEEE}
}

@article{hu2022tackling,
  title={Tackling long code search with splitting, encoding, and aggregating},
  author={Hu, Fan and Wang, Yanlin and Du, Lun and Zhang, Hongyu and Han, Shi and Zhang, Dongmei and Li, Xirong},
  journal={arXiv preprint arXiv:2208.11271},
  year={2022}
}

@inproceedings{leclair2019neural,
  title={A neural model for generating natural language summaries of program subroutines},
  author={LeClair, Alexander and Jiang, Siyuan and McMillan, Collin},
  booktitle={2019 IEEE/ACM 41st International Conference on Software Engineering (ICSE)},
  pages={795--806},
  year={2019},
  organization={IEEE}
}

@article{zhu2019automatic,
  title={Automatic code summarization: A systematic literature review},
  author={Zhu, Yuxiang and Pan, Minxue},
  journal={arXiv preprint arXiv:1909.04352},
  year={2019}
}

@inproceedings{haiduc2010supporting,
  title={Supporting program comprehension with source code summarization},
  author={Haiduc, Sonia and Aponte, Jairo and Marcus, Andrian},
  booktitle={Proceedings of the 32nd ACM/IEEE International Conference on Software Engineering-Volume 2},
  pages={223--226},
  year={2010}
}

@inproceedings{moreno2013automatic,
  title={Automatic generation of natural language summaries for java classes},
  author={Moreno, Laura and Aponte, Jairo and Sridhara, Giriprasad and Marcus, Andrian and Pollock, Lori and Vijay-Shanker, K},
  booktitle={2013 21st International conference on program comprehension (ICPC)},
  pages={23--32},
  year={2013},
  organization={IEEE}
}

@inproceedings{eddy2013evaluating,
  title={Evaluating source code summarization techniques: Replication and expansion},
  author={Eddy, Brian P and Robinson, Jeffrey A and Kraft, Nicholas A and Carver, Jeffrey C},
  booktitle={2013 21st International Conference on Program Comprehension (ICPC)},
  pages={13--22},
  year={2013},
  organization={IEEE}
}

@inproceedings{iyer2016summarizing,
  title={Summarizing source code using a neural attention model},
  author={Iyer, Srinivasan and Konstas, Ioannis and Cheung, Alvin and Zettlemoyer, Luke},
  booktitle={54th Annual Meeting of the Association for Computational Linguistics 2016},
  pages={2073--2083},
  year={2016},
  organization={Association for Computational Linguistics}
}

@article{fang2024esale,
  title={Esale: Enhancing code-summary alignment learning for source code summarization},
  author={Fang, Chunrong and Sun, Weisong and Chen, Yuchen and Chen, Xiao and Wei, Zhao and Zhang, Quanjun and You, Yudu and Luo, Bin and Liu, Yang and Chen, Zhenyu},
  journal={IEEE Transactions on Software Engineering},
  volume={50},
  number={8},
  pages={2077--2095},
  year={2024},
  publisher={IEEE}
}

@article{sun2024extractive,
  title={An extractive-and-abstractive framework for source code summarization},
  author={Sun, Weisong and Fang, Chunrong and Chen, Yuchen and Zhang, Quanjun and Tao, Guanhong and You, Yudu and Han, Tingxu and Ge, Yifei and Hu, Yuling and Luo, Bin and others},
  journal={ACM Transactions on Software Engineering and Methodology},
  volume={33},
  number={3},
  pages={1--39},
  year={2024},
  publisher={ACM New York, NY, USA}
}

@article{fried2022incoder,
  title={Incoder: A generative model for code infilling and synthesis},
  author={Fried, Daniel and Aghajanyan, Armen and Lin, Jessy and Wang, Sida and Wallace, Eric and Shi, Freda and Zhong, Ruiqi and Yih, Wen-tau and Zettlemoyer, Luke and Lewis, Mike},
  journal={arXiv preprint arXiv:2204.05999},
  year={2022}
}

@article{su2024distilled,
  title={Distilled GPT for source code summarization},
  author={Su, Chia-Yi and McMillan, Collin},
  journal={Automated Software Engineering},
  volume={31},
  number={1},
  pages={22},
  year={2024},
  publisher={Springer}
}

@inproceedings{bansal2021project,
  title={Project-level encoding for neural source code summarization of subroutines},
  author={Bansal, Aakash and Haque, Sakib and McMillan, Collin},
  booktitle={2021 IEEE/ACM 29th International Conference on Program Comprehension (ICPC)},
  pages={253--264},
  year={2021},
  organization={IEEE}
}

@article{allamanis2017learning,
  title={Learning to represent programs with graphs},
  author={Allamanis, Miltiadis and Brockschmidt, Marc and Khademi, Mahmoud},
  journal={arXiv preprint arXiv:1711.00740},
  year={2017}
}

@article{guo2020graphcodebert,
  title={Graphcodebert: Pre-training code representations with data flow},
  author={Guo, Daya and Ren, Shuo and Lu, Shuai and Feng, Zhangyin and Tang, Duyu and Liu, Shujie and Zhou, Long and Duan, Nan and Svyatkovskiy, Alexey and Fu, Shengyu and others},
  journal={arXiv preprint arXiv:2009.08366},
  year={2020}
}

@article{ye2023cp,
  title={Cp-bcs: Binary code summarization guided by control flow graph and pseudo code},
  author={Ye, Tong and Wu, Lingfei and Ma, Tengfei and Zhang, Xuhong and Du, Yangkai and Liu, Peiyu and Ji, Shouling and Wang, Wenhai},
  journal={arXiv preprint arXiv:2310.16853},
  year={2023}
}

@article{zhang2024review,
  title={A review of automatic source code summarization},
  author={Zhang, Xuejun and Hou, Xia and Qiao, Xiuming and Song, Wenfeng},
  journal={Empirical Software Engineering},
  volume={29},
  number={6},
  pages={162},
  year={2024},
  publisher={Springer}
}

@article{lin2021semantic,
  title={Semantic feature learning via dual sequences for defect prediction},
  author={Lin, Junhao and Lu, Lu},
  journal={IEEE Access},
  volume={9},
  pages={13112--13124},
  year={2021},
  publisher={IEEE}
}

@inproceedings{wan2018improving,
  title={Improving automatic source code summarization via deep reinforcement learning},
  author={Wan, Yao and Zhao, Zhou and Yang, Min and Xu, Guandong and Ying, Haochao and Wu, Jian and Yu, Philip S},
  booktitle={Proceedings of the 33rd ACM/IEEE international conference on automated software engineering},
  pages={397--407},
  year={2018}
}

@article{vaswani2017attention,
  title={Attention is all you need},
  author={Vaswani, Ashish and Shazeer, Noam and Parmar, Niki and Uszkoreit, Jakob and Jones, Llion and Gomez, Aidan N and Kaiser, {\L}ukasz and Polosukhin, Illia},
  journal={Advances in neural information processing systems},
  volume={30},
  year={2017}
}

@article{wang2021codet5,
  title={Codet5: Identifier-aware unified pre-trained encoder-decoder models for code understanding and generation},
  author={Wang, Yue and Wang, Weishi and Joty, Shafiq and Hoi, Steven CH},
  journal={arXiv preprint arXiv:2109.00859},
  year={2021}
}

@article{roziere2023code,
  title={Code llama: Open foundation models for code},
  author={Roziere, Baptiste and Gehring, Jonas and Gloeckle, Fabian and Sootla, Sten and Gat, Itai and Tan, Xiaoqing Ellen and Adi, Yossi and Liu, Jingyu and Sauvestre, Romain and Remez, Tal and others},
  journal={arXiv preprint arXiv:2308.12950},
  year={2023}
}

@inproceedings{zhang2020retrieval,
  title={Retrieval-based neural source code summarization},
  author={Zhang, Jian and Wang, Xu and Zhang, Hongyu and Sun, Hailong and Liu, Xudong},
  booktitle={Proceedings of the ACM/IEEE 42nd International Conference on Software Engineering},
  pages={1385--1397},
  year={2020}
}

@article{alon2018code2seq,
  title={code2seq: Generating sequences from structured representations of code},
  author={Alon, Uri and Brody, Shaked and Levy, Omer and Yahav, Eran},
  journal={arXiv preprint arXiv:1808.01400},
  year={2018}
}

@inproceedings{qiu2024software,
  title={Software Defect Prediction Based on Double Traversal AST},
  author={Qiu, Shaoming and Bicong, E and Huang, Xinchen and Liu, Liangyu},
  booktitle={2024 8th Asian Conference on Artificial Intelligence Technology (ACAIT)},
  pages={1665--1674},
  year={2024},
  organization={IEEE}
}

@inproceedings{kovalenko2019pathminer,
  title={Pathminer: a library for mining of path-based representations of code},
  author={Kovalenko, Vladimir and Bogomolov, Egor and Bryksin, Timofey and Bacchelli, Alberto},
  booktitle={2019 IEEE/ACM 16th International Conference on Mining Software Repositories (MSR)},
  pages={13--17},
  year={2019},
  organization={IEEE}
}

@inproceedings{ahmed2024automatic,
  title={Automatic semantic augmentation of language model prompts (for code summarization)},
  author={Ahmed, Toufique and Pai, Kunal Suresh and Devanbu, Premkumar and Barr, Earl},
  booktitle={Proceedings of the IEEE/ACM 46th international conference on software engineering},
  pages={1--13},
  year={2024}
}

@article{li2023starcoder,
  title={Starcoder: may the source be with you!},
  author={Li, Raymond and Allal, Loubna Ben and Zi, Yangtian and Muennighoff, Niklas and Kocetkov, Denis and Mou, Chenghao and Marone, Marc and Akiki, Christopher and Li, Jia and Chim, Jenny and others},
  journal={arXiv preprint arXiv:2305.06161},
  year={2023}
}

@inproceedings{lomshakov2024proconsul,
  title={Proconsul: Project context for code summarization with llms},
  author={Lomshakov, Vadim and Podivilov, Andrey and Savin, Sergey and Baryshnikov, Oleg and Lisevych, Alena and Nikolenko, Sergey},
  booktitle={Proceedings of the 2024 Conference on Empirical Methods in Natural Language Processing: Industry Track},
  pages={866--880},
  year={2024}
}

@inproceedings{ahmed2022few,
  title={Few-shot training llms for project-specific code-summarization},
  author={Ahmed, Toufique and Devanbu, Premkumar},
  booktitle={Proceedings of the 37th IEEE/ACM international conference on automated software engineering},
  pages={1--5},
  year={2022}
}
\end{document}